\documentclass[aps,prl,showpacs,amssymb,superscriptaddress,nofootinbib,twocolumn]{revtex4-1} 
\usepackage{graphicx}
\usepackage{amsmath}
\usepackage{color}
\usepackage{cancel}

\newcommand{\ket}[1]{\ensuremath{\left|{#1}\right\rangle}}

\newcommand{\braket}[2]{\ensuremath{\langle{#1}|{#2}\rangle}}
\newcommand{\brakket}[3]{\ensuremath{\langle{#1}|{#2}|{#3}\rangle}}

\newcommand{\da}[0]{\dagger}

\begin{document}

\title{Short-time quantum detection: probing quantum fluctuations}

\author{Marco del Rey}
\affiliation{Instituto de F\'{\i}sica Fundamental, CSIC,
  Serrano 113-B, 28006 Madrid, Spain}
\email{marco.del.rey@iff.csic.es}

\author{Carlos Sab\'in}
\affiliation{Instituto de F\'{\i}sica Fundamental, CSIC,
  Serrano 113-B, 28006 Madrid, Spain}

\author{Juan Le\'on}
\affiliation{Instituto de F\'{\i}sica Fundamental, CSIC,
  Serrano 113-B, 28006 Madrid, Spain}

\date{\today}

\begin{abstract}
In this work we study the information provided by a detector click on the state of an initially excited two level system. By computing the time evolution of the corresponding conditioned probability beyond the rotating wave approximation, as needed for short time analysis, we show that a click in the detector is related with the decay of the source only for long times of interaction. For short times, non-rotating wave approximation effects, like self-excitations of the detector, forbid a na\"{i}ve interpretation of the detector readings. These effects might appear in circuit QED experiments.
\end{abstract}


\maketitle

Quantum detection theory was created to study the behavior of detectors in presence of radiation \cite{Glauber}. Highly satisfactory up to date, it relies on the conspicuous rotating wave approximation (RWA), which neglects the so-called counterrotating terms, irrelevant in most cases. These terms give important contributions for strong atom-field couplings and very short times as compared to the system time scale, meaning that for any effect beyond RWA to be acknowledged our measurements must be very precise and the observation times quite fast. This is particularly problematic for Quantum Optics experiments, due to the very small matter-radiation coupling and the fact that observation times must be at the femtosecond scale for most cases (nanosecond for hyperfine qubits), which is ridiculously small for current experiments ($\sim \mu s$  for trapped ions  \cite{Blatt}). 

However, circuit QED \cite{reviewnature} provides a framework in which those phenomena are accessible to explore. By using superconducting qubits as artificial atoms coupled to a transmission line, one gets a system which behaves analogously as a 1-D  radiation-matter interaction in Quantum Optics, while working at the microwave frequency range \cite{clarke08}. In this setup, parameters can be easily tuned, and the coupling between qubits and transmission line can be modulated up to ultrastrong levels. \cite{devoretultrastrong, ultrastrong2} Besides, fast qubit state readout ($\sim ns$) is possible using a pulsed DC-SQUID scheme \cite{measurementqubits1} . 
 
In those conditions, phenomena beyond RWA have already been reported \cite{niemczyk10}, \cite{forn-diaz10}. In this regime  Glauber's theory is no longer valid and quantum detectors should be described by  a non-RWA model. 

A counterintuitive direct consequence of the breakdown of the RWA is that a detector in its ground state interacting with the vacuum of the field has a certain probability of getting excited and emitting a photon. There is  however not a real consensus on the physical reality of this effect. Introducing counterrotating terms is interpreted by some to be a problem as the processes described by those terms seem virtual. It seems counterintuitive to accept that a detector in a ground state in the vacuum could get excited.  As a matter of fact, there have been attempts of suggesting effective detector models by imposing this phenomenon to be impossible \cite{piazzacosta}. 

We should however point out here that there is no problem whatsoever with energy conservation, as unitary evolution of the states under a hermitian hamiltonian directly guarantees it. Any peculiar effect at that respect is linked to the fact that the initial state considered has not  definite energy, since the state "detector and field in their ground states" is not an eigenstate of the full Hamiltonian beyond RWA. 

In this work we would rather step out of this discussion and by taking the theoretical models coming from circuit QED, without imposing any additional constraints, we will study the following setup:  a source $S$ initially excited, a detector $D$ initially in the ground state and both interacting with the electromagnetic field in its vacuum state.  If the detector clicks at a given time, does it mean that the source is now in the ground state? This problem amounts to compute the probability of decay of the source, conditioned to the excitation of the detector. We will show that, unlike Glauber's RWA detector in which this conditioned probability would be equal to 1 at any time, this circuit QED  detector only achieves this value at long times, due to the impact of non-RWA effects.

More precisely, our model has two superconducting qubits, $S$ and $D$, with two levels $g$ and $e$ and separated a distance $r$. Let us consider an initial moment $t=0$ where $S$ is excited, $D$ (which represents the detector) is in its ground state and there are no excitations in the transmission line, which will be open, so enabling a continuum of modes. Representing the states in terms of  qubit ($S$, $D$) and field ($F$) free eigenstates with the notation $\ket{\psi}=\ket{SDF}$, the initial state would be $\ket{i}_{t=0}=\ket{eg0}$. 

After a certain time $t$, if  we measure qubit $D$  and it results excited, that would na\"{i}vely lead us to think $S$ has decayed and produced a photon which  has then later been absorbed by $D$. We intend to proof otherwise by quantifying what information about the state of $S$  can be extracted by knowing qubit $D$ state after a certain time $t$. For that we will compute  the probability $\mathcal{P}_{S_g/D_e} (t)$ of $S$ to have decayed at a certain instant $t$, conditioned we have measured $D$ excited at that same moment: 
\begin{equation}
\mathcal{P}_{S_g/D_e} (t)=\frac{\mathcal{P}_{[ge*]}}{\mathcal{P}_{[*e*]}} = \frac{\sum_{F} |\brakket{geF}{e^{-iHt/\hbar}}{eg0}|^2}{\sum_{n,F} |\brakket{neF}{e^{-iHt/\hbar}}{eg0}|^2}
\label{eq:conditionedprobability}
\end{equation}
$\mathcal{P}_{[ge*]}$ being the probability of $S$ being in the ground state and $D$ excited and $\mathcal{P}_{[*e*]}$ the total probability of excitation of $D$.

From here on, we will consider the following Hamiltonian \cite{yurkedenker, guillejuanjokike}: 
\begin{align}
H&=H_0+H_I,\nonumber\\
H_0&=\sum_{A=\{S,D\}}\frac{\hbar \Omega_A}{2} \sigma^{A}_z +\int_{-\infty}^\infty dk  \hbar \omega_k a^\da_k a_k,\nonumber\\
H_I&=-\sum_{A=\{S,D\}} d_A V(x_A)\sigma_x^{A} ,
\label{eq:totalhamiltonian}
\end{align}
 Here  $x_A$ corresponds to the position of the qubit $A$, $\hbar \Omega_A$ is the gap between levels for qubit $A$ and $V$ refers to the 1-dimensional field which expands as: 
\begin{equation}
V (x) = i \int_{-\infty}^{\infty} dk \sqrt{N\omega_k} e^{ikx}  a_{k}+ H.c.,
\label{eq:onedimensionalfield}
\end{equation}
This field has a continuum of Fock operators $[a_k,a^{\dag}_{k'}]=\delta(k-k'),$ and a linear spectrum, $\omega_k =
v|k|$, where $v$ is the propagation velocity of the field. The normalization and the speed of photons, $v=(cl)^{-1/2},$ depend on the microscopic details such as the capacitance and inductance per unit length, $c$ and $l.$ 
Note that this model, a dimensional reduction of the matter- radiation hamiltonian with two-level atoms and the electromagnetic field, is formally equivalent to the Unruh- de Witt model. \cite{Unruh}. 
For our calculations, we will make use of the interaction picture, so we let the initial state $|eg0\rangle$ evolve for a lapse of time $t$ as: 
\begin{align}
&|\psi(t)\rangle =  U_I(t) \ket{ e g 0}=\mathcal{T} \{ e^{-i\int_0^t dt' H_I(t')/\hbar}\}\ket{ e g 0}\nonumber\\
&=I \ket{ e g 0}+X\ket{ g e 0}+\sum_k A_{1,k}\ket{ g g 1_k}+\sum_k  B_{1,k}\ket{ e e 1_k} \nonumber\\
&\quad +\sum_{kk'} A_{2,kk'} \ket{ e g 2_{kk'}} + \sum_{kk'} B_{2,kk'} \ket{ g e 2_{kk'}}+...,\label{eq:evolution}
\end{align}
Here and in the following we will only make explicit the terms that contain contributions for the probabilities up to $d_A^4$. For example,  terms with 3 or more photons in the amplitude will be excluded, as they give contributions of $\mathcal{O}(d_A^6)$. 

Having $\mathcal{M}(t;nF)=\braket{n e F}{\psi(t)}$, the probabilities needed for the computation of $\mathcal{P}_{S_g/D_e} (t)$ (\ref{eq:conditionedprobability}) can be written down using (\ref{eq:evolution}) as:
\begin{align}
\mathcal{P}_{[ g e *]}  &= \sum_{F} |\brakket{g e F}{U_I(t)}{e g 0}|^2=  \sum_{F} |\mathcal{M}(t;gF)|^2 \nonumber\\
&= |X|^2+\sum |B_2|^2+ \dots\label{eq:probs2}\\
\mathcal{P}_{[* e *]} 	&=\sum_{n,F}|\brakket{n e F}{U_I(t)}{e g 0}|^2= \sum_{n,F}|\mathcal{M}(t;nF)|^2 \nonumber\\
&= |X|^2+\sum |B_1|^2+\sum |B_2|^2+\dots
\label{eq:probs3}
\end{align}
The first building block needed is $|X|^2$. Note that:
\begin{equation}
\mathcal{P}_{[g e 0]}  =|\brakket{g e 0}{U_I(t)}{e g 0}|^2=|\mathcal{M}(t;g0)|^2 =|X|^2\label{eq:probs1}
\end{equation}
To evaluate $|X|^2$ up to fourth order in perturbation theory, one must consider X has no contributions neither for orders 0 or 1, so the calculation must be performed for orders 2 and above. As a fact, order 2 alone will suffice. This calculation has been already performed in the appendix of \cite{conjuanjo}, where the dimensionless coupling $
K_A=\frac{4d_A^2N}{\hbar^2v}$, with $\ A=\{S,D\}$, acts as the perturbative parameter. We will restrict to times where $K_A\omega_At \ll 1 $, where our perturbative approach remains valid .

\begin{figure}[h]
\begin{center}
\includegraphics[width=0.45\textwidth]{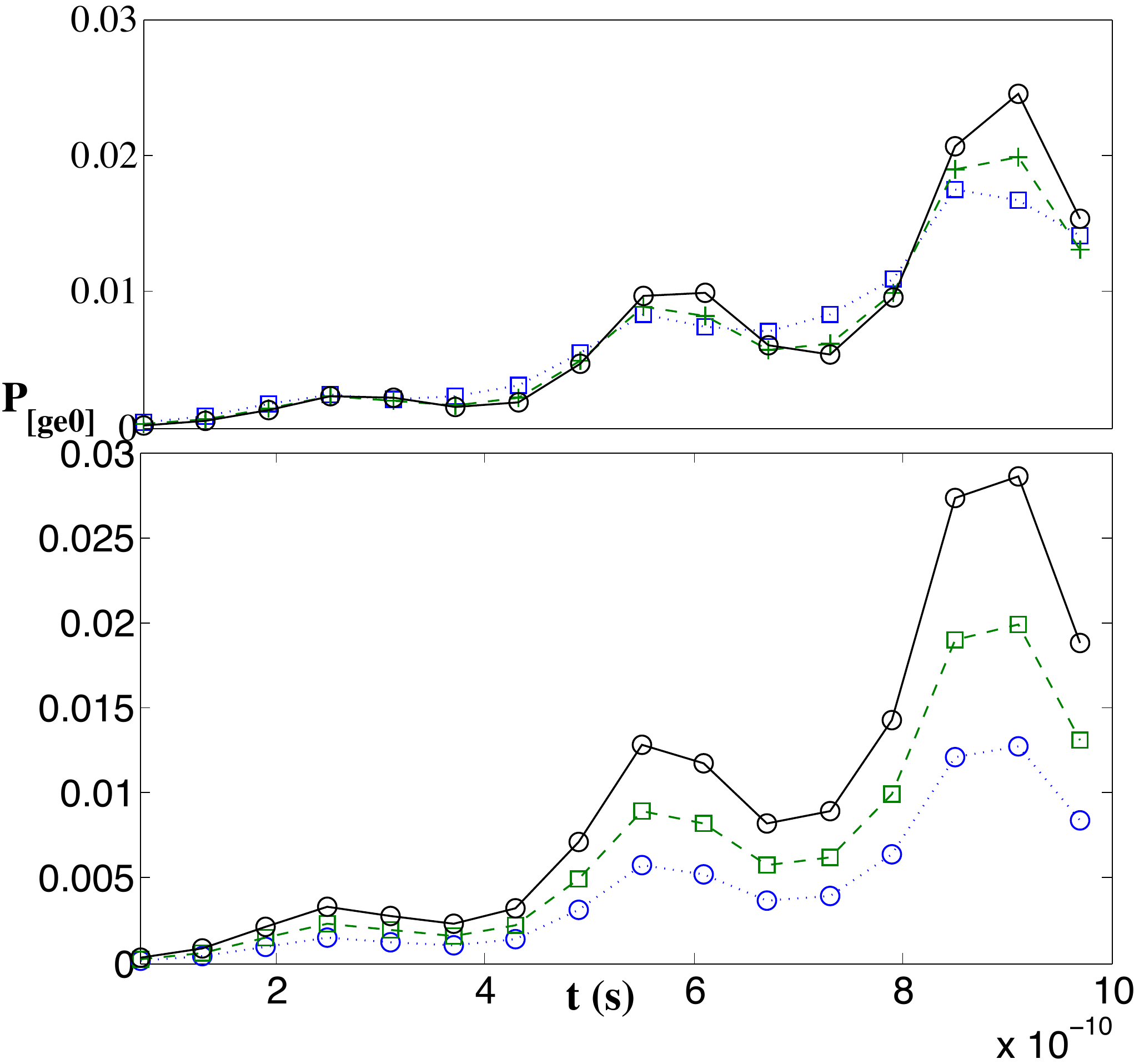}
\end{center}
\caption{a) $P_{[eg0]}$  in front of $t$ in $s$ for three different values of the distance between qubits $2\pi\frac{r}{\lambda}=0.1$ (dotted, squares, blue),  $0.3$ (dashed, crosses, green) and  $0.5$ (solid, circles, black).For all cases the coupling strength is $K_S=K_D=7.5 \cdot 10^{-3}$ and the qubit frequency  $\Omega/(2\pi)= 1\, Ghz$. b)  $P_{[eg0]}$  in front of $t$ in $s$ for three different values of the coupling strength $K=K_S=K_D=6\cdot 10^{-3}$  (dotted, squares, blue),  $7.5 \cdot 10^{-3}$
  (dashed, crosses, green) and  $9 \cdot 10^{-3}$ (solid, circles, black) . For all cases $2\pi\frac{r}{\lambda}=0.3$ and $\Omega/(2\pi)= 1\, GHz$. } 
\label{fig:1}
\end{figure}

In Fig.  \ref{fig:1}  we sketch the evolution of the probability $\mathcal P_{[ge0]}$ with time, and its dependence with the coupling and the distance between qubits. Typical values for couplings and distances for a setup in circuit QED are considered from here on.  At these early stages $\mathcal P_{[ge0]}$ is highly oscillatory in time. For a given time, the probability always grows with the coupling strength but depends of  the distance in different ways.
To proceed with the calculation of  $\mathcal P_{[ge*]}$, the term $B_2$ must be evaluated. As the final bare state associated with this term has two photons, this implies automatically that orders 0 and 1 are discarded. Once again, order 2 alone fits. 
Once calculated, it must be squared and summed, splitting into two terms, a ``direct'' one, just the product of the square of the emission amplitudes, and a ``crossed'' one, which looks like a photon exchange. The summation of the direct terms implies the appearance of  expected divergences which can be resolved using a regularization procedure as the one sketched in the appendix of \cite{conjuan}. This procedure requires the times of analysis to be larger than a certain cutoff time $t_0$, which  in this case is related with the typical size of a superconducting qubit $d\simeq10^{-5} \,m$ \cite{blais} and the propagation velocity of the field quanta: $v\simeq 10^{8}\,m/s$. Thus, $t_0=d/c\simeq1=10^{-13}\,s$ far below the times considered in this work. A detailed treatment of the procedure related with the emission probabilities at short times will be published elsewhere.

\begin{figure}[h]
\begin{center}
\includegraphics[width=0.45\textwidth]{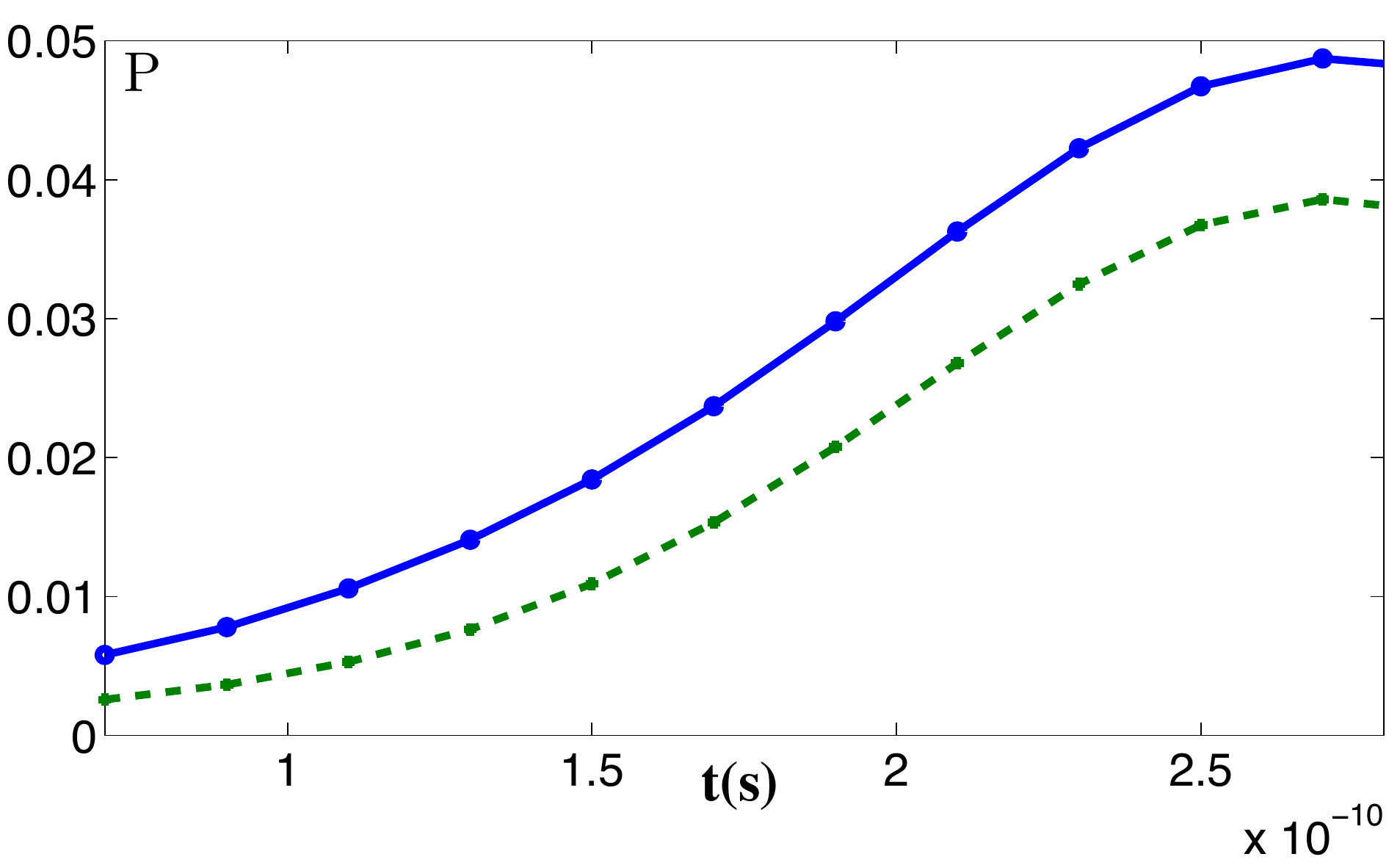}
\end{center}
\caption{a) $P_{[eg*]}$ (solid, blue, circles) and $P_{[eg0]}$ (dashed, green, crosses) in front of $t$ in $s$ with a distance $2\pi\frac{r}{\lambda}=0.5$,  a coupling strength of $K=K_S=K_D=3 \cdot 10^{-2}$ and a qubit frequency of $\Omega/(2\pi)= 1\, GHz$ ($\Omega=\Omega_S=\Omega_D$). The difference between the two graphs is the non-RWA term $\sum |B_2|^2$.} 
\label{fig:2}
\end{figure}

Notice that $B_2$ is only non-zero beyond the RWA.  In Fig. \ref{fig:2} we compare  $\mathcal P_{[ge*]}$ with  $\mathcal P_{[ge0]}$ and the impact of this non-RWA contribution  is seen in the sub-nanoseconds regime for a large value of the coupling strength. At larger times, the impact diminishes and $\mathcal P_{[ge*]}\simeq\mathcal P_{[ge0]}$.

\begin{figure}[h]
\begin{center}
\includegraphics[width=0.46\textwidth]{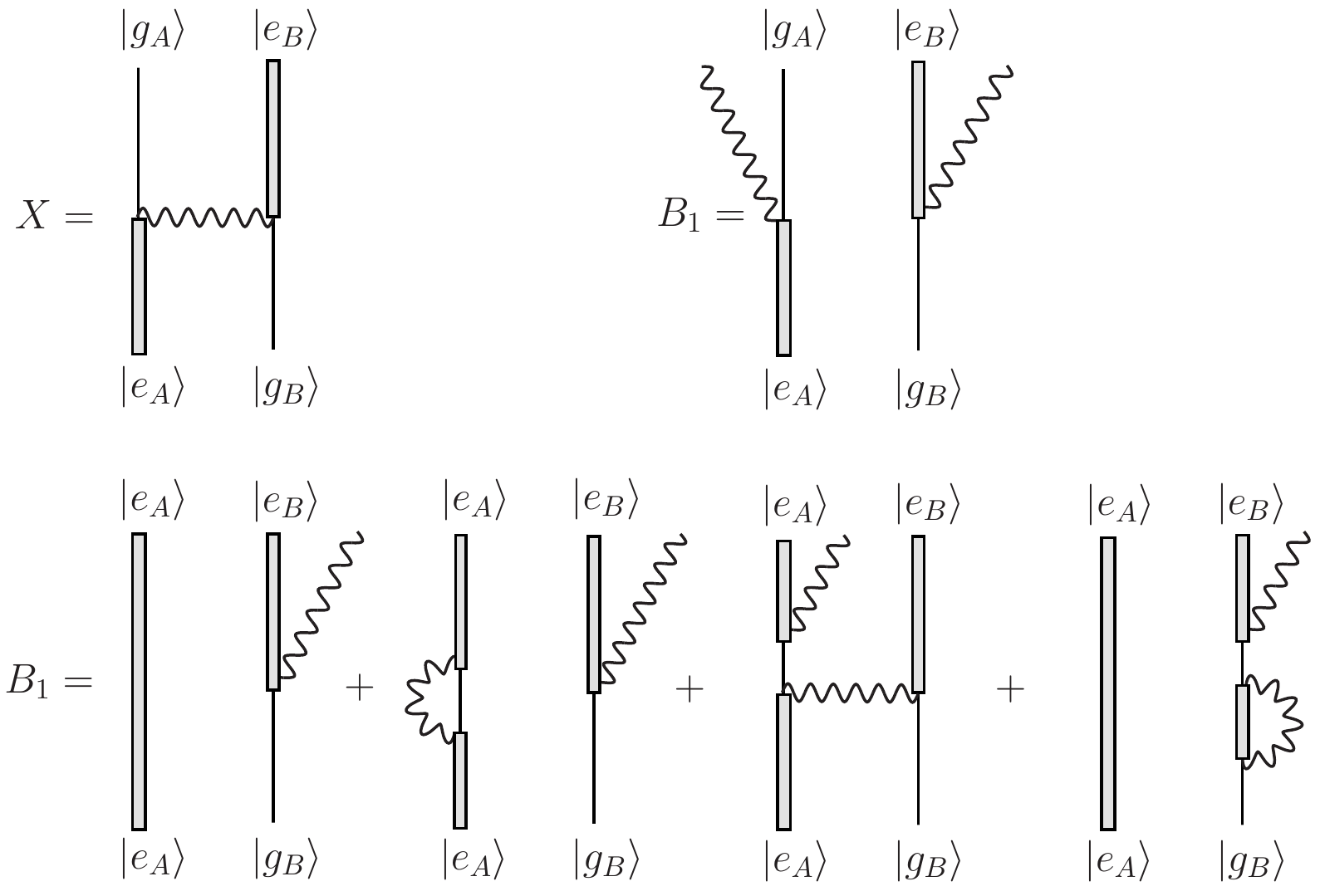}
\end{center}
\caption{The diagrams contributing to $X$, $B_1$ and $B_2$. $X$ represents the amplitude for photon exchange between source and detector, while $B_2$ is just the amplitude for two single photon emissions, one at each qubit. The leading order contribution to $B_1$ is the amplitude for a single photon emission at the detector qubit (fourth diagram), but third-order one-loop corrections (first two) and a photon exchange accompanied by an emission at the source have to be also taken into account. $B_1$ and $B_2$ are completely non-RWA diagrams.
} 
\label{fig:3}
\end{figure}

The last probability of interest, $\mathcal P_{[*e*]}$, needs $\sum|B_1|^2$, which is again a completely non-RWA contribution. For that case the situation gets more complicated, as there are interfering processes of orders 1 and 3 leading to that final state. This calculation has been the focus of a recent work \cite{nuestrofermi}. 
The four diagrams contributing to $\sum|B_1|^2$ up to fourth order in perturbation theory can be seen in fig (\ref{fig:3}). The leading order contribution is just the probability of self-excitation of the detector, (fourth diagram in fig. \ref{fig:3}) and the other contributions come from the interference of this diagram with the other three. In particular, interference with the third diagram is crucial for causality \cite{nuestrofermi}.

\begin{figure}[h!]
\begin{center}
\includegraphics[width=0.45\textwidth]{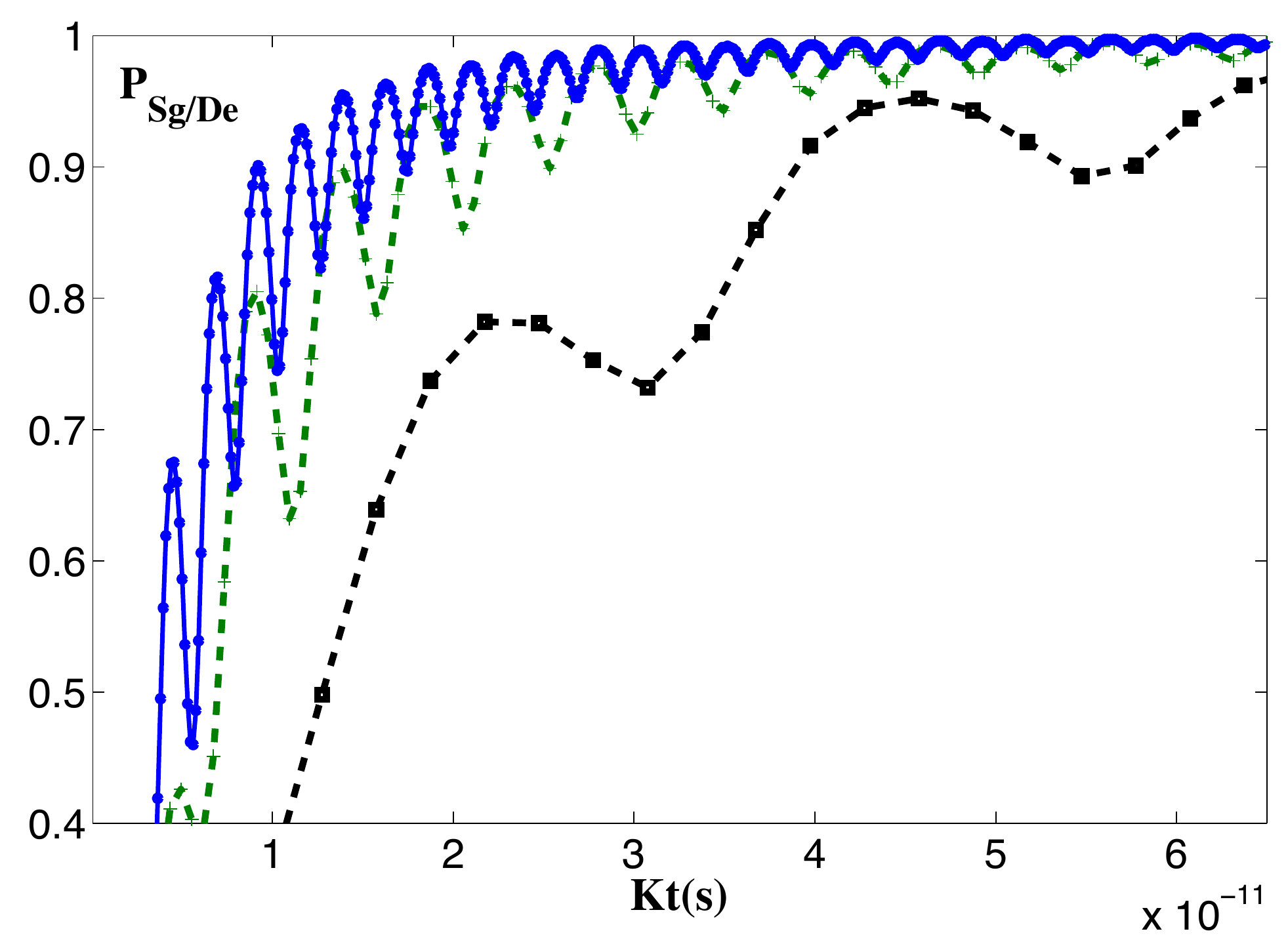}
\end{center}
\caption{$\mathcal{P}_{S_g/D_e} (t)$ (\ref{eq:conditionedprobability}) in front of $Kt$ for three different values of the  coupling strength of $K=\,K_S= K_D=7.5\cdot 10^{-3}$ (solid, blue, circles), $1.5\, 10^{-2}$(dashed, green, crosses), $7.5 \cdot 10^{-2}$  (dashed, black, squares). In the three cases  $2\pi\frac{r}{\lambda}= 1$ and  $\Omega/(2\pi) = 1\, GHz$ ($\Omega=\Omega_S=\Omega_D$).  } 
\label{fig:4}
\end{figure}

\begin{figure}[h!]
\begin{center}
\includegraphics[width=0.45\textwidth]{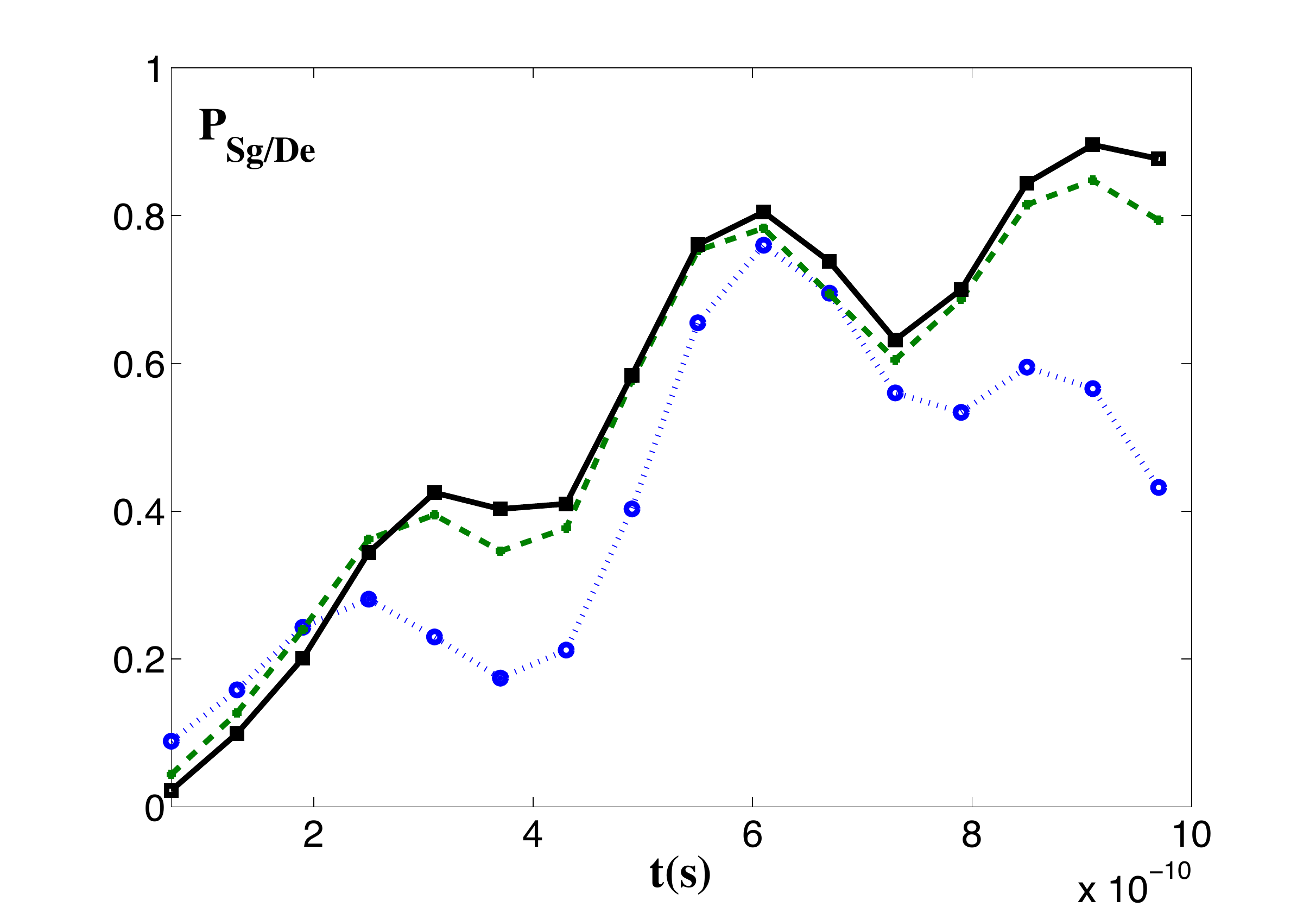}
\end{center}
\caption{$\mathcal{P}_{S_g/D_e} (t)$ (\ref{eq:conditionedprobability}) in front of $t$ in $s$ for three different values of the distance $2\pi\frac{r}{\lambda}=0.5$ (dotted, blue, circles), $0.75$ (dashed, green, crosses), $1$ (solid, black, squares). In the three cases the  coupling strength $K=\,K_S= K_D=1.5\cdot 10^{-2}$  and $\Omega/(2\pi)=1\, GHz$ ($\Omega=\Omega_S=\Omega_D$).} 
\label{fig:5}
\end{figure}

With the previous probabilities computed we can finally address the conditioned probability  $\mathcal{P}_{S_g/D_e} (t)$, which can be calculated as shown in (\ref{eq:conditionedprobability}). Note that in the RWA $\mathcal{P}_{S_g/D_e} (t)=1$ at any time, since $\mathcal{P}_{[*e*]}=\mathcal{P}_{[ge*]}=\mathcal{P}_{[ge0]}$.  We have seen however that non-RWA contributions to $\mathcal{P}_{[*e*]}$ have a sizable impact in the sub-nanosecond regime. The effect of these contributions to the evolution of  $\mathcal{P}_{S_g/D_e} (t)$ can be seen in Figs. \ref{fig:4} and \ref{fig:5}, where the consequences of changing the coupling and the distance between qubits are considered. The first thing we notice in fig. \ref{fig:4} is that for short times the information provided by the detector is not very much related to the state of the source, that is, self-excitations and other non-RWA phenomena dominate over the photon exchange between source and detector. For the cases considered, only at interaction times $t \gtrsim 1\,ns	\simeq  1/\Omega$  the conditioned probability converges to the RWA prediction, that is, the excitation of the detector is a reliable way to detect the decay of the source.  Since the non-RWA contributions are more relevant for large couplings and short distances, the convergence is faster as the distance grows and the couplings diminish, as can be seen in Figs. \ref{fig:4} and \ref{fig:5}.

The above theoretical results could have an impact in real experiments of circuit QED. In particular, a typical setup to measure the internal state of a flux qubit coupled to a transmission line consists of a SQUID surrounding the qubit. Although the total measurement process could take up to tens of nanoseconds, most of the time the coupling SQUID-qubit is much stronger than $K$, \cite{measurementqubits1} and the dynamics qubit-transmission line is effectively frozen. Thus this dynamics is only important during the activation of the SQUID, a process that may be in the nanosecond regime. Therefore, if at $t=0$ the system is in the initial state of $\ket{eg0}$ (see \cite{nuestrofermi}) for the preparation of such a state) and then we measure the state of the qubit, our results show that a click of the detector could be as related to a self-excitation as to an absorption of a photon emitted by the source.

To conclude, we have considered a system of two superconducting qubits coupled to an open transmission line, which can be suitably described in the framework of 1-D QED with two-level (artificial) atoms. By using perturbation theory in the interaction picture we have computed at short times the probability of excitation of a detector qubit in the presence of an originally excited source qubit subjected to a possible decay.   Our main result is that for typical circuit QED parameters, a significative amount of time is needed before we start trusting the detector results as informative with respect to the source.  This is an effect of the breakdown of the RWA in circuit QED, which gives rise to non-RWA phenomena like self-excitations. By neglecting the counterrotating terms, the sophisticated reality of the short-time phenomena stays hidden and, in particular, a total reliability on the information coming out of the detector is wrongly derived for all time-scales. It is noteworthy to mention that this result applies in general to other setups and detectors, although for extremely short times, and it is in the case of circuit QED where it might affect the interpretation of coming experimental results.
 
This work is supported by Spanish MICINN Project FIS2008-05705 and CAM research consortium QUITEMAD S2009-ESP-1594. M. del Rey acknowledges the support from CSIC JAE-PREDOC2010 grant and Residencia de Estudiantes.

\bibliographystyle{apsrev}
\bibliography{references}

\begin{thebibliography}{17}
\expandafter\ifx\csname natexlab\endcsname\relax\def\natexlab#1{#1}\fi
\expandafter\ifx\csname bibnamefont\endcsname\relax
  \def\bibnamefont#1{#1}\fi
\expandafter\ifx\csname bibfnamefont\endcsname\relax
  \def\bibfnamefont#1{#1}\fi
\expandafter\ifx\csname citenamefont\endcsname\relax
  \def\citenamefont#1{#1}\fi
\expandafter\ifx\csname url\endcsname\relax
  \def\url#1{\texttt{#1}}\fi
\expandafter\ifx\csname urlprefix\endcsname\relax\def\urlprefix{URL }\fi
\providecommand{\bibinfo}[2]{#2}
\providecommand{\eprint}[2][]{\url{#2}}

\bibitem[{\citenamefont{Glauber}(1963)}]{Glauber}
\bibinfo{author}{\bibfnamefont{R.~J.} \bibnamefont{Glauber}},
  \bibinfo{journal}{Phys. Rev.} \textbf{\bibinfo{volume}{130}},
  \bibinfo{pages}{2529} (\bibinfo{year}{1963}).

\bibitem[{\citenamefont{Leibfried et~al.}(2003)\citenamefont{Leibfried, Blatt,
  Monroe, and Wineland}}]{Blatt}
\bibinfo{author}{\bibfnamefont{D.}~\bibnamefont{Leibfried}},
  \bibinfo{author}{\bibfnamefont{R.}~\bibnamefont{Blatt}},
  \bibinfo{author}{\bibfnamefont{C.}~\bibnamefont{Monroe}}, \bibnamefont{and}
  \bibinfo{author}{\bibfnamefont{D.}~\bibnamefont{Wineland}},
  \bibinfo{journal}{Rev. Mod. Phys.} \textbf{\bibinfo{volume}{75}},
  \bibinfo{pages}{281} (\bibinfo{year}{2003}).

\bibitem[{\citenamefont{You and Nori}(2011)}]{reviewnature}
\bibinfo{author}{\bibfnamefont{J.~Q.} \bibnamefont{You}} \bibnamefont{and}
  \bibinfo{author}{\bibfnamefont{F.}~\bibnamefont{Nori}},
  \bibinfo{journal}{Nature} \textbf{\bibinfo{volume}{474}},
  \bibinfo{pages}{589} (\bibinfo{year}{2011}).

\bibitem[{\citenamefont{{Clarke} and {Wilhelm}}(2008)}]{clarke08}
\bibinfo{author}{\bibfnamefont{J.}~\bibnamefont{{Clarke}}} \bibnamefont{and}
  \bibinfo{author}{\bibfnamefont{F.~K.} \bibnamefont{{Wilhelm}}},
  \bibinfo{journal}{Nature} \textbf{\bibinfo{volume}{453}},
  \bibinfo{pages}{1031} (\bibinfo{year}{2008}).

\bibitem[{\citenamefont{{Devoret} et~al.}(2007)\citenamefont{{Devoret},
  {Girvin}, and {Schoelkopf}}}]{devoretultrastrong}
\bibinfo{author}{\bibfnamefont{M.~H.} \bibnamefont{{Devoret}}},
  \bibinfo{author}{\bibfnamefont{S.}~\bibnamefont{{Girvin}}}, \bibnamefont{and}
  \bibinfo{author}{\bibfnamefont{R.}~\bibnamefont{{Schoelkopf}}},
  \bibinfo{journal}{Annalen der Physik} \textbf{\bibinfo{volume}{519}},
  \bibinfo{pages}{767} (\bibinfo{year}{2007}).

\bibitem[{\citenamefont{Bourassa et~al.}(2009)\citenamefont{Bourassa, Gambetta,
  Abdumalikov, Astafiev, Nakamura, and Blais}}]{ultrastrong2}
\bibinfo{author}{\bibfnamefont{J.}~\bibnamefont{Bourassa}},
  \bibinfo{author}{\bibfnamefont{J.~M.} \bibnamefont{Gambetta}},
  \bibinfo{author}{\bibfnamefont{A.~A.} \bibnamefont{Abdumalikov}},
  \bibinfo{author}{\bibfnamefont{O.}~\bibnamefont{Astafiev}},
  \bibinfo{author}{\bibfnamefont{Y.}~\bibnamefont{Nakamura}}, \bibnamefont{and}
  \bibinfo{author}{\bibfnamefont{A.}~\bibnamefont{Blais}},
  \bibinfo{journal}{Phys. Rev. A} \textbf{\bibinfo{volume}{80}},
  \bibinfo{pages}{032109} (\bibinfo{year}{2009}).

\bibitem[{\citenamefont{Picot et~al.}(2010)\citenamefont{Picot, Schouten,
  Harmans, and Mooij}}]{measurementqubits1}
\bibinfo{author}{\bibfnamefont{T.}~\bibnamefont{Picot}},
  \bibinfo{author}{\bibfnamefont{R.}~\bibnamefont{Schouten}},
  \bibinfo{author}{\bibfnamefont{C.~J. P.~M.} \bibnamefont{Harmans}},
  \bibnamefont{and} \bibinfo{author}{\bibfnamefont{J.~E.} \bibnamefont{Mooij}},
  \bibinfo{journal}{Phys. Rev. Lett.} \textbf{\bibinfo{volume}{105}},
  \bibinfo{pages}{040506} (\bibinfo{year}{2010}).

\bibitem[{\citenamefont{{Niemczyk} et~al.}(2010)\citenamefont{{Niemczyk},
  {Deppe}, {Huebl}, {Menzel}, {Hocke}, {Schwarz}, {Garcia-Ripoll}, {Zueco},
  {H{\"u}mmer}, {Solano} et~al.}}]{niemczyk10}
\bibinfo{author}{\bibfnamefont{T.}~\bibnamefont{{Niemczyk}}},
  \bibinfo{author}{\bibfnamefont{F.}~\bibnamefont{{Deppe}}},
  \bibinfo{author}{\bibfnamefont{H.}~\bibnamefont{{Huebl}}},
  \bibinfo{author}{\bibfnamefont{E.~P.} \bibnamefont{{Menzel}}},
  \bibinfo{author}{\bibfnamefont{F.}~\bibnamefont{{Hocke}}},
  \bibinfo{author}{\bibfnamefont{M.~J.} \bibnamefont{{Schwarz}}},
  \bibinfo{author}{\bibfnamefont{J.~J.} \bibnamefont{{Garcia-Ripoll}}},
  \bibinfo{author}{\bibfnamefont{D.}~\bibnamefont{{Zueco}}},
  \bibinfo{author}{\bibfnamefont{T.}~\bibnamefont{{H{\"u}mmer}}},
  \bibinfo{author}{\bibfnamefont{E.}~\bibnamefont{{Solano}}},
  \bibnamefont{et~al.}, \bibinfo{journal}{Nature Physics}
  \textbf{\bibinfo{volume}{6}}, \bibinfo{pages}{772} (\bibinfo{year}{2010}).

\bibitem[{\citenamefont{{Forn-D{\'{\i}}az}
  et~al.}(2010)\citenamefont{{Forn-D{\'{\i}}az}, {Lisenfeld}, {Marcos},
  {Garc{\'{\i}}a-Ripoll}, {Solano}, {Harmans}, and {Mooij}}}]{forn-diaz10}
\bibinfo{author}{\bibfnamefont{P.}~\bibnamefont{{Forn-D{\'{\i}}az}}},
  \bibinfo{author}{\bibfnamefont{J.}~\bibnamefont{{Lisenfeld}}},
  \bibinfo{author}{\bibfnamefont{D.}~\bibnamefont{{Marcos}}},
  \bibinfo{author}{\bibfnamefont{J.~J.} \bibnamefont{{Garc{\'{\i}}a-Ripoll}}},
  \bibinfo{author}{\bibfnamefont{E.}~\bibnamefont{{Solano}}},
  \bibinfo{author}{\bibfnamefont{C.~J.~P.~M.} \bibnamefont{{Harmans}}},
  \bibnamefont{and} \bibinfo{author}{\bibfnamefont{J.~E.}
  \bibnamefont{{Mooij}}}, \bibinfo{journal}{Phys. Rev. Lett.}
  \textbf{\bibinfo{volume}{135}}, \bibinfo{pages}{237001}
  (\bibinfo{year}{2010}).

\bibitem[{\citenamefont{{Costa} and {Piazza}}(2009)}]{piazzacosta}
\bibinfo{author}{\bibfnamefont{F.}~\bibnamefont{{Costa}}} \bibnamefont{and}
  \bibinfo{author}{\bibfnamefont{F.}~\bibnamefont{{Piazza}}},
  \bibinfo{journal}{New Journal of Physics} \textbf{\bibinfo{volume}{11}},
  \bibinfo{pages}{113006} (\bibinfo{year}{2009}), \eprint{0805.0806}.

\bibitem[{\citenamefont{{Yurke} and {Denker}}(1984)}]{yurkedenker}
\bibinfo{author}{\bibfnamefont{B.}~\bibnamefont{{Yurke}}} \bibnamefont{and}
  \bibinfo{author}{\bibfnamefont{J.~S.} \bibnamefont{{Denker}}},
  \bibinfo{journal}{Phys. Rev. A} \textbf{\bibinfo{volume}{29}},
  \bibinfo{pages}{1419} (\bibinfo{year}{1984}).

\bibitem[{\citenamefont{{Romero} et~al.}(2009)\citenamefont{{Romero}, {Jos{\'e}
  Garc{\'{\i}}a-Ripoll}, and {Solano}}}]{guillejuanjokike}
\bibinfo{author}{\bibfnamefont{G.}~\bibnamefont{{Romero}}},
  \bibinfo{author}{\bibfnamefont{J.}~\bibnamefont{{Jos{\'e}
  Garc{\'{\i}}a-Ripoll}}}, \bibnamefont{and}
  \bibinfo{author}{\bibfnamefont{E.}~\bibnamefont{{Solano}}},
  \bibinfo{journal}{Physica Scripta Volume T} \textbf{\bibinfo{volume}{137}},
  \bibinfo{pages}{014004} (\bibinfo{year}{2009}).

\bibitem[{\citenamefont{Unruh}(1976)}]{Unruh}
\bibinfo{author}{\bibfnamefont{W.~G.} \bibnamefont{Unruh}},
  \bibinfo{journal}{Phys. Rev. D} \textbf{\bibinfo{volume}{14}},
  \bibinfo{pages}{870} (\bibinfo{year}{1976}).

\bibitem[{\citenamefont{Sab\'\i{}n et~al.}(2010)\citenamefont{Sab\'\i{}n,
  Garc\'\i{}a-Ripoll, Solano, and Le\'on}}]{conjuanjo}
\bibinfo{author}{\bibfnamefont{C.}~\bibnamefont{Sab\'\i{}n}},
  \bibinfo{author}{\bibfnamefont{J.~J.} \bibnamefont{Garc\'\i{}a-Ripoll}},
  \bibinfo{author}{\bibfnamefont{E.}~\bibnamefont{Solano}}, \bibnamefont{and}
  \bibinfo{author}{\bibfnamefont{J.}~\bibnamefont{Le\'on}},
  \bibinfo{journal}{Phys. Rev. B} \textbf{\bibinfo{volume}{81}},
  \bibinfo{pages}{184501} (\bibinfo{year}{2010}).

\bibitem[{\citenamefont{Le\'on and Sab\'\i{}n}(2009)}]{conjuan}
\bibinfo{author}{\bibfnamefont{J.}~\bibnamefont{Le\'on}} \bibnamefont{and}
  \bibinfo{author}{\bibfnamefont{C.}~\bibnamefont{Sab\'\i{}n}},
  \bibinfo{journal}{Phys. Rev. A} \textbf{\bibinfo{volume}{79}},
  \bibinfo{pages}{012304} (\bibinfo{year}{2009}).

\bibitem[{\citenamefont{Blais et~al.}(2004)\citenamefont{Blais, Huang,
  Wallraff, Girvin, and Schoelkopf}}]{blais}
\bibinfo{author}{\bibfnamefont{A.}~\bibnamefont{Blais}},
  \bibinfo{author}{\bibfnamefont{R.-S.} \bibnamefont{Huang}},
  \bibinfo{author}{\bibfnamefont{A.}~\bibnamefont{Wallraff}},
  \bibinfo{author}{\bibfnamefont{S.~M.} \bibnamefont{Girvin}},
  \bibnamefont{and} \bibinfo{author}{\bibfnamefont{R.~J.}
  \bibnamefont{Schoelkopf}}, \bibinfo{journal}{Phys. Rev. A}
  \textbf{\bibinfo{volume}{69}}, \bibinfo{pages}{062320}
  (\bibinfo{year}{2004}).

\bibitem[{\citenamefont{{Sabin} et~al.}(2011)\citenamefont{{Sabin}, {del Rey},
  {Garcia-Ripoll}, and {Leon}}}]{nuestrofermi}
\bibinfo{author}{\bibfnamefont{C.}~\bibnamefont{{Sabin}}},
  \bibinfo{author}{\bibfnamefont{M.}~\bibnamefont{{del Rey}}},
  \bibinfo{author}{\bibfnamefont{J.~J.} \bibnamefont{{Garcia-Ripoll}}},
  \bibnamefont{and} \bibinfo{author}{\bibfnamefont{J.}~\bibnamefont{{Leon}}},
  \bibinfo{journal}{ArXiv}  (\bibinfo{year}{2011}), \eprint{1103.4129}.

\end{thebibliography}

\end{document}